# Multi-scale characterization of hexagonal Si-4H: a hierarchical nanostructured material

Silvia Pandolfi,[1*] Shiteng Zhao[2-3©],, John Turner[2], Peter Ercius[2], Chengyu Song[2], Rohan Dhall[2], Nicolas Menguy[1], Yann Le Godec[1], Alexandre Courac[1,4], Andrew M. Minor[2-3], Jon Eggert[5] and Leora E. Dresselhaus-Marais[5#]

[1]*Sorbonne Université, Muséum National d'Histoire Naturelle, UMR CNRS 7590, Institut de Minéralogie, de Physique des Matériaux et de Cosmochimie, IMPMC, 75005 Paris, France*
[2]*National Center for Electron Microscopy, Molecular Foundry, Lawrence Berkeley National Laboratory, Berkeley, CA, USA*
[3]*Department of Materials Science and Engineering, University of California, Berkeley, CA, USA*
[4]*Institut universitaire de France, IUF, 75005 Paris, France*
[5]*Lawrence Livermore National Laboratory. 7000 East Avenue, Livermore, California 94550, USA*
[*]now at: *SLAC National Accelerator Laboratory, 2575 Sand Hill Road, Menlo Park, California 94025, USA*
[©] now at: *Professor, School of Materials Science and Engineering, Beihang University, Beijing,100191. China*
[#]now at: *Assistant Professor of Materials Science & Engineering at Stanford University, Stanford, CA, USA*

In this work we present a detailed structural characterization of Si-4H, a newly discovered bulk form of hexagonal silicon (Si) with potential optoelectronic applications. Using multi-scale imaging, we reveal a hierarchical structure in the morphology of Si-4H obtained from high-pressure synthesis. We demonstrate discrete structural units, *platelets*, at an intermediate length-scale between the bulk pellets synthesized at high pressures and the flake-like crystallites inferred in previous studies. Direct observation of the platelets reveals their 2D structure, with planar faces spanning hundreds of nanometers to a few micrometers and thicknesses of only tens of nanometers. We separated and dispersed small packets of quasi-single platelets, which enabled us to analyze the crystalline domains within each grain. With this view, we demonstrate that Si-4H platelets represent the smallest crystalline structural units, which can bend at the single-domain level. Our characterization of the quasi-2D, flexible platelets of hexagonal Si-4H and proof of concept that the platelets can be dispersed and manipulated quite simply demonstrate opportunities to design novel optoelectronic and solar devices.

## I. INTRODUCTION

Progress in optoelectronic technology requires new materials whose properties can be tailored as needed [1–4]. Silicon (Si) currently dominates the semiconductor industry [5,6] as it is abundant and inexpensive, and it can be produced with remarkably high purity, which ensures the high carrier lifetimes necessary to maximize its efficiency [7]. Nevertheless, Si's efficiency in optoelectronics and photovoltaics is limited by its indirect bandgap [8,9]. There is, therefore, significant motivation to design new Si-based materials that can optimize Si's interaction with light and easily integrate them into current technology. Towards this goal, a plethora of new Si-based materials have been proposed, including low-density allotropes [1–3,10–12], low-dimensional systems [13–23] and metastable phases recoverable from high-pressure (HP) [24–32].

Hexagonal Si has emerged as a promising candidate for photovoltaic applications [33–36], as it is predicted to shift from an indirect- to a direct-bandgap semiconductor with the application of strain [34]. Proof of concept for the synthesis of hexagonal Si has been reported from chemical methods [24], as an inclusion in polycrystalline Si nanostructures [37–40], from HP treatment [25,26,41,42], from epitaxial growth on a GaP nanowire (NW) core [43] and in GaP/Si/SiGe NWs [44]. Beyond synthesis, limited characterization work has been performed to connect hexagonal Si's properties back to its predicted industrial utility, as synthesis of the pure material has only recently become available [29,45]. We recently established the first synthesis of pure hexagonal Si from HP treatment via annealing of metastable Si-III, and identified its crystal structure [29]: Si-4H polytype, with ABCB stacking[1]. Alternatively, Si-4H can also be obtained via annealing of metastable $Si_{24}$, a procedure that earlier this year led to the synthesis of polycrystalline grains with ~0.5 μm single-crystal domains [45]. Upon HP synthesis, we demonstrated substantial nanostructuring, and inferred that Si-4H is comprised of flake-like crystallites set by the high density of stacking faults along the $c$ axis, i.e., the stacking direction of the hexagonal planes. The highly textured nanostructure presents questions as to how the nanostructure relates to the powder-like bulk material and how this could affect Si-4H's optoelctronic properties. On one side, quantum confinement and the presence of crystalline defects might limit the lifetime of the free carriers, while on the other it could modify the bangdap value and increase the efficiency of radiative recombination thanks to carrier concentration and indirect space delocalization [46]. The discrepancy between the 1.65 eV emission measured in nanostructured Si-4H [29] compared to the 1.2 eV absorption onset recently measured in polycrystalline Si-4H with ~0.5μm grain size suggests that the texture and crystal size strongly influence the optoelectronic properties of hexagonal Si. A detailed knowledge of Si-4H nanostructure is thus needed to better understand its impact on Si-4H properties and technological applications.

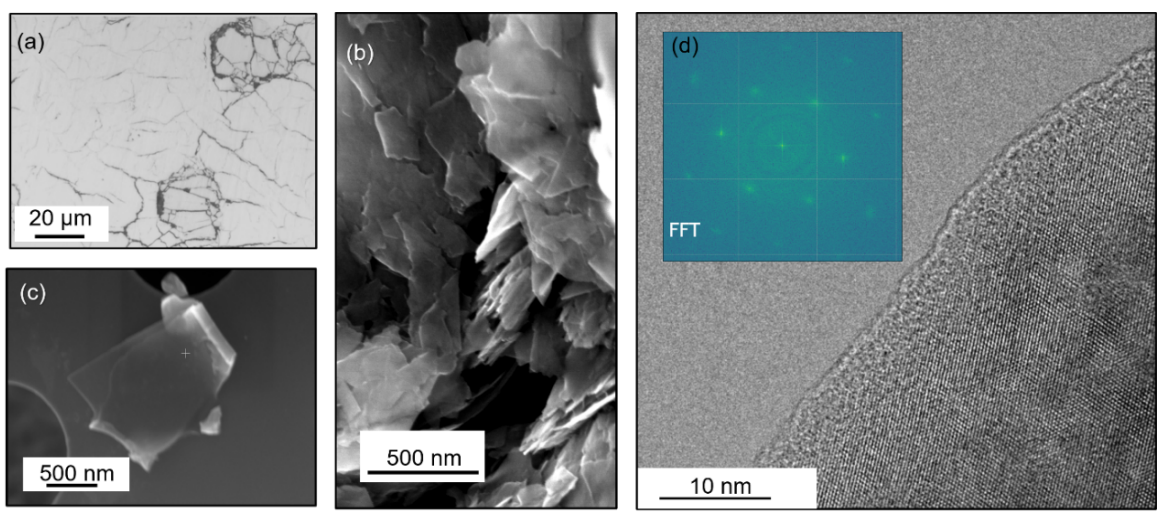

**FIG. 1.** Multi-scale analysis of the hierarchical structure of Si-4H. A combination of optical and electron microscopy techniques was used to access the different length-scales. (a) Optical microscope image of the cracks observed in the polished sample surface. (b) SEM image of the pristine Si-4H morphology. We observe quasi-2D discrete structural units, that we refer to as *platelets*. (c) SEM image of quasi-single platelets dispersed on a TEM grid. (d) TEM image of a single platelet. The FFT pattern confirms the single crystal nature of the platelet imaged here.

The work presented here takes a direct look at the microstructure of Si-4H from HP synthesis and at the key structural information required to evaluate its applications. Towards that goal, we provide detailed characterization of Si-4H's structure, revealing a hierarchical nanostructure that offers unique opportunities for future engineering-based studies. Our multi-scale approach demonstrates discrete 2D crystalline units, that we refer to as *platelets*, that link the highly textured nanostructure and the powder-like bulk structure (Fig. 1). We exploit the weak interaction forces between platelets to disperse and characterize single grains. With our single-platelet analysis, we observe single-crystal domains with very low angle boundaries to accommodate bending, with only localized oxidation at the platelets' edges. In light of these new findings, Si-4H platelets present new opportunities for functional applications, e.g. thin coatings for multi-junction solar cells or flexible components for microelectronic devices.

---

[1] Hexagonal Si can assume different crystal structures, called polytypes, that differ only in the stacking of the hexagonal planes along the $c$ axis.

## II. DETAILED METHODOLOGY

A. **Si-4H Synthesis Procedure.** The sample we describe in this paper was synthesized at the European Synchrotron Radiation Facility (ESRF) on the ID06-LVP beamline [47], using a Deformation-DIA multianvil press and a 10/5 assembly. The sample was a compacted pellet (1.95 mm diameter, 2.12 mm height and prepared in an Ar-filled glovebox) of high-purity Si (Alfa Aesar). The sample was surrounded by an *h*BN sleeve (2.75 mm outer diameter (OD), 1.96 mm inner diameter (ID), 2.12 mm height) and a Ta foil (25 μm thickness, 2.90 mm OD) resistive furnace. The Ta foil was inserted into a 10 mm octahedron edge length (OEL) $Cr_2O_3$-doped MgO octahedral pressure medium, along with $ZrO_2$ plugs. The sample was compressed up to ~13 GPa and then heated up to ~870 K while maintaining the pressure quasi-constant. After quenching, a pure Si-II (i.e. the metallic HP phase) was stabilized. We then released the pressure, transforming the HP Si-II phase into BC8 cubic Si-III phase, which is metastable at ambient conditions. We then formed hexagonal Si-4H by heating the recovered Si-III pellet to ~200°C under vacuum (~$10^{-3}$ mbar).

B. **Sample preparation for electron microscopy.** The sample was polished for surface characterization using diamond-based mechanical polishing paper with 1 μm and 0.1 μm particle grit. We confirmed the surface finish via optical microscopy, and evaluated that the flat polished surface had pervasive cracks and voids due to underlying grain structure. Beyond the surface polishing, electron-transparent samples suitable for Transmission Electron Microscopy (TEM) were prepared by cratching the polished surface with 1-μm grit polishing paper. The expelled Si-4H particles were then collected with a drop of ethanol and dispersed onto a TEM grid after sonicating for 1-2 minutes. We used a TEM copper grid (300 mesh) with Quantifoil® support film (orthogonal array of 1.2 μm holes on a 20 nm thick film).

C. **Structural characterization: multi-scale microscopy.** Structural characterization was performed at the National Center for Electron Microscopy in the Molecular Foundry at Lawrence Berkeley National Laboratory. Surface validation was performed with a Leica DFC295 optical microscope using objectives with ≤100X magnification. Scanning Electron Microscopy (SEM) was performed using the FEI Helios G4 UX Focused Ion Beam system. Ultra-High Resolution (UHR) was achieved in the so-called *immersion mode*, with the sample immersed in the lenses' field. HR-TEM images were acquired on a Tecnai F20 with 200 keV acceleration voltage and 0.2 nm Scherzer resolution. Electron Energy Loss Spectroscopy (EELS) experiment was carried out on a Gatan Tridiem GIF camera, with 1 nm probe size and 0.5 eV energy resolution. Probe convergent angle is 10 mrad. GIF collection angle is 42 mrad.

## III. EXPERIMENTAL RESULTS

The surface structures were characterized by comparing the polished and unpolished surfaces of the as-synthesized pellet (Fig. 1(a-b)). Our previous work suggested ordered "nanoflakes" from TEM measurements and bulk "powder" pellets from X-ray diffraction [29]. The new findings from surface characterization suggested grain structures at an intermediate lengthscale not observed previously (Supplemental Material, Sec. 1). We split this section into two parts to resolve the structure at this scale fully. We first show the SEM results that resolve the structural units on the unpolished pellet surfaces, demonstrating that the synthesis forms discrete 2D grains (Sec. III-A , Fig. 1(b)). Secondly, we describe how we then separated and dispersed the Si-4H grains (Fig. 1(c)) to study them in further detail by TEM. Analysis of the crystalline domain combining TEM imaging, Fast-Fourier Transform (FFT) analysis and EELS measurements enabled us to resolve the size distribution of the single-crystalline domains once dispersed and evaluate their internal mosaicity (Sec. III-B, Fig. 1(d)).

**III-A. Discrete Structural Units: Si-4H platelets**

A network of fine cracks was revealed in the polished Si-4H's surface (Fig. 1(a)) whose internal structure suggests the presence of discrete grains (Supplemental Material, Sec.1). To explore the structure of sub-µm scale grains in the as-grown samples, we zoomed into the unpolished surfaces of Si-4H using the SEM's Ultra-High-Resolution (UHR) mode. Representative images of the unpolished surfaces are shown in Fig. 2(a), demonstrating large populations of grains with sub-µm dimensions that are packed into a range of orientations. The observed grains form apparently planar structures, as we illustrate with green arrows showing the flat faces and yellow arrows showing the thin dimension. In this work, we refer to these planar structures as *platelets*, as they have strong similarities to the platelet-like crystals of SiC-4H [48].

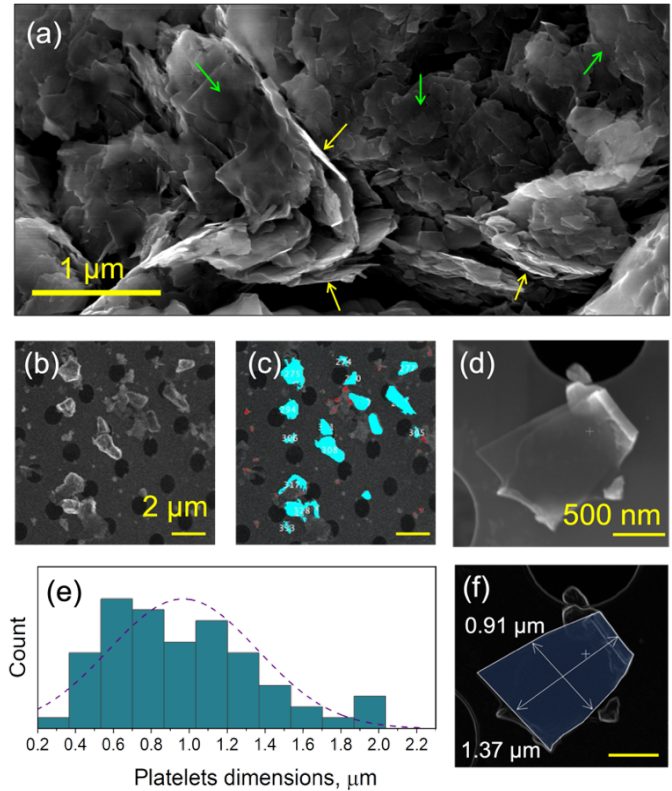

**FIG. 2.** UHR-SEM images and size analysis of Si-4H platelets. (a) Group of particles on the unpolished surface of the sample. The image shows both very thin particles' faces (yellow arrows) and bigger, flat ones (green ones), demonstrating their anisotropic shape. (b,d) Dispersed platelets on a TEM grid prepared according to the procedure described in Sec. II-B. The light blue overlays in (c) show the domains identified in (b) by edge and particle analysis. In (f) the boundaries of (d) were consensed manually, as the platelets were superposed. (e) Histogram of the platelets' sizes.

The platelets in Fig. 2 appear loosely packed, preferentially along their flat faces, suggesting only weak binding forces connecting the platelets. Following the hypothesis of weak inter-platelet binding, we developed a dispersion procedure (Sec. II-B) to manipulate the grains. We then used the dispersed platelets to measure the size distribution of the platelets for comparison to the nanoflakes inferred in our previous work. Separation of the platelets required only gentle sonication, confirming that the platelets bind weakly to each other along the wide faces. Representative images of dispersed platelets deposited on a TEM grid and a detailed analysis of the size distributions are shown in Fig. 2(b-f).

Fig. 2(b,d) shows representative regions of the dispersed platelets; the platelets deposited either as single units or small collections that pack preferentially parallel to their wide faces. We quantify the particles' sizes with Sobel edge-detection and an intensity threshold to reconstruct and identify each object. We demonstrate the effectiveness of these segmentation results for the isolated platelets in Fig. 2b with the overlaid regions in Fig. 2(c). When the platelets are not isolated, i.e., there is superposition (Fig 2(d)) and/or the platelets are in contact with their neighbors, we manually selected the appropriate boundaries to discriminate between discrete grains in the agglomerate (Fig 2(f)). The results from our size analysis are reported in the histogram in Fig 2(e). The size distribution of the platelets is centered around 1µm, with a calculated FWHM of ~400 nm. We note that the platelets size has a very wide variance, ranging from 200 nm to 2 µm. Since platelets of different sizes are observed also in the "as-grown" pellet (Fig.2(a)), we conclude that the variance in size is intrinsic to the synthesis and not degradation from the dispersion procedure (e.g., fracture of large platelets).

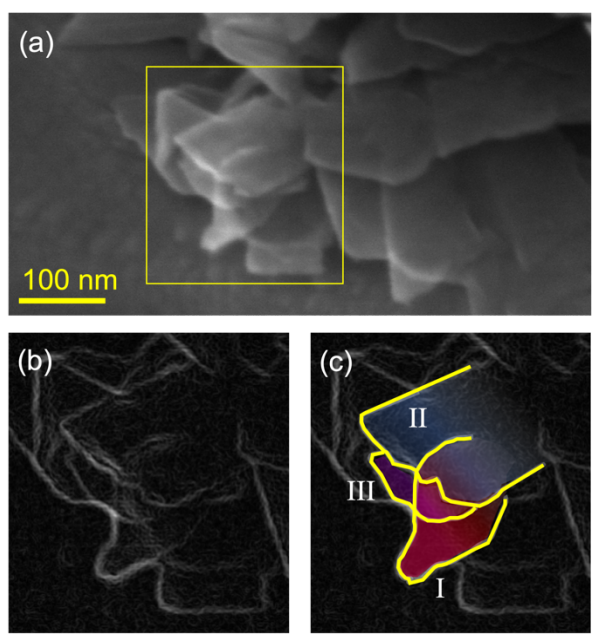

**FIG. 3.** (a) UHR-SEM images of platelets' stack. The raw results from Sobel edge detection method on the area highlighted in yellow are shown in (b). We identify different particles with bold yellow outlines and different colored overlayers in (c).

In contrast to their wide faces, the thickness of the Si-4H platelets was more difficult to measure directly because of its thinness. The subset of grains in Fig. 3(a) allowed us to quantitatively constrain the thickness indirectly. Fig. 3(b) shows white lines that outline specific platelet domains, as evaluated from the Sobel edge-detection result for the yellow boxed region in Fig. 3(a). The continuity, curvature and corners between the lines for each grain demonstrate that the raw boundaries extracted from Fig. 3(b) must form three distinct platelets, as indicated by yellow outlines and shaded overlays for each grain in Fig 3(c). For the stack of platelets identified in Fig. 3(c), the edges of the red platelet (I) are clearly visible, even when they cross the area that corresponds to the blue (II) and purple (III) ones. The fact that we see the outlines of all three platelets over the span of the platelet I's face implies that we must be seeing through at least one of them, no matter which one is the top of the stack. For example, if I is on top, it must be transparent, as we still observe the outline of the platelets II and III. Following the outline analysis, we conclude that the platelets are partially transparent to the SEM electron beam, setting a ~100-nm upper limit for the platelet height[2].

### III-B. Crystallinity within Si-4H Platelets

For a deeper understanding of the platelets structure and crystallinity, we zoom into the structure of single grains by measuring the dispersed platelets with high-resolution (HR) TEM. At this magnification, we characterize the local structure within *each* platelet to resolve the crystalline and amorphous domains and their extent. To analyze the extent and orientation of the crystalline domains within each platelet, we used Fourier analysis (details in Supplemental Material, Sec.3). This approach uses post-processing to reconstruct a map of the orientation of each crystalline region within the sample[3]. Fig. 4 shows the results of Fourier analysis on a representative TEM image of Si-4H platelets.

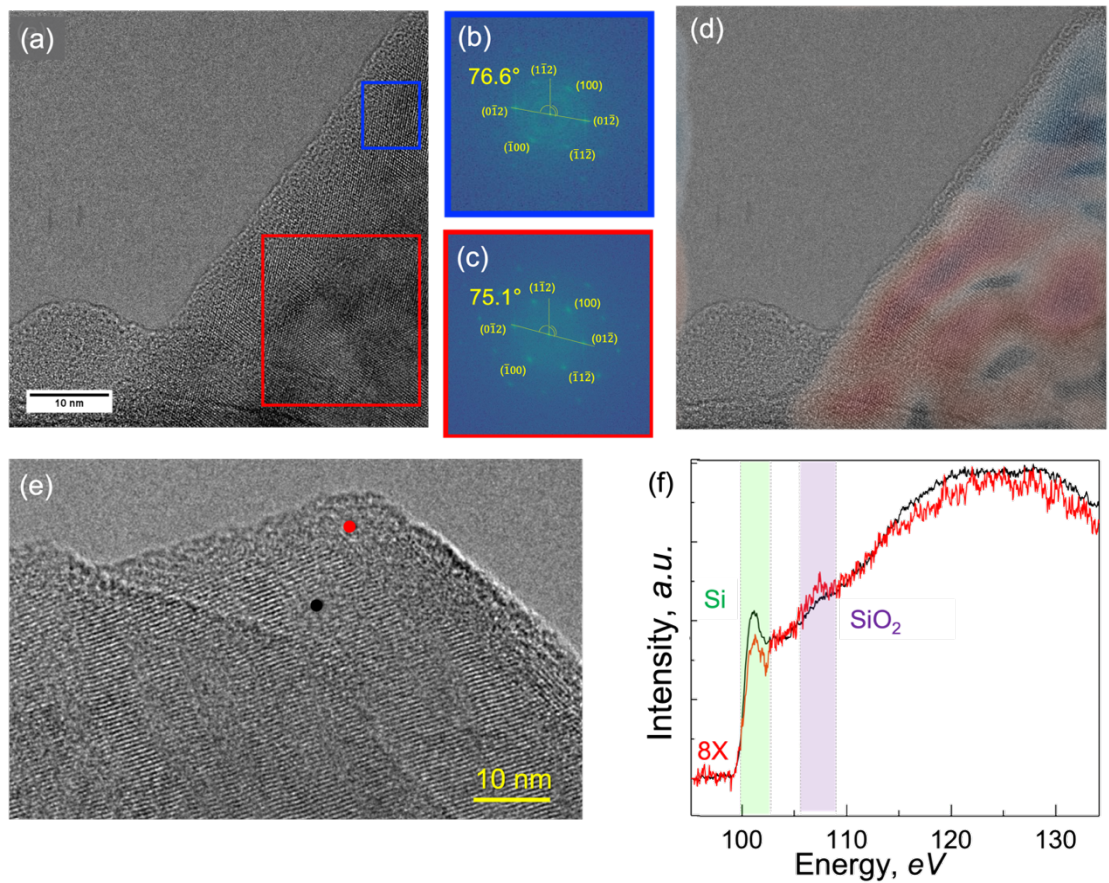

---

[2] The SEM was operated using a 5kV acceleration voltage. We note that this conclusion also assumes that the discrete units located with our analysis were indeed single platelets. It is possible that, for example, each of the analyzed regions could actually correspond to a stack of platelets. For that case, our analysis would conclude that the *full stack* of platelets that comprise the red one has at most ~100nm thickness.

[3] We note that this approach is only feasible for very thin samples, which is satisfied for single- and few- platelets of only ~10s nanometers thickness.

**FIG. 4.** TEM-based analysis of single Si-4H platelets. (a) HR-TEM image. (b),(c) FFT-TEM pattern of selected area, highlighted in (a) with red and blue rectangles, respectively. The peaks are indexed, and the angles between the [$0\bar{1}2$] reciprocal direction and the vertical direction of the image are indicated. (d) Inverse filtered FFT images; the overlays highlight the area corresponding to each crystalline orientation. (e) HR-TEM image of suspended platelets. (f) EELS spectra from two characteristic points marked in (e) by black and red dots, respectively. The red spectrum was multiplied by a factor 8 for comparison to the black curve, as the sample thinned significantly close to its edge. The regions highlighted in green and purple indicate the signature of Si and $SiO_2$, respectively.

The TEM image in Fig. 4(a) shows two representative Si-4H platelets; the platelet on the right side of the image exhibit a good cristallinity, while the ones on the left of the image seem to be amorphous showing no evidence of periodicity. Indexation of FFT patterns related to different regions of the platelets, i.e., the ones highlighted in blue and red in Fig. 4(a) confirms a common [021] zone axis for the two different domains, with however a slight rotation ~1.5° with respect to each other (Fig. 4(b-c)). We perform Fourier filtering on the full FFT-TEM image, demonstrating that the two Si-4H domains cover the full surface of the platelet on the right, as shown by the blue and red overlays on the original TEM image in Fig. 4(d). More details on the Fourier analysis procedure and results can be found in the Supplemental Material, Sec. 3. We note that the small tilt between the two Si-4H crystalline domains is consistent with a bend in the originally single-crystalline region getting accommodated with a very low-angle bundary. This suggests that the platelets, suspended over the holes in the TEM substrate, may accommodate strain via formation of defects, such as low-angle grain boundaries that enable them to flex at the single-platelet scale.

In classical Si solar cells, the efficiency is minimized by a passivating layer of oxidation at the surface. Given the very thin Si-4H platelet dimensions, we combine TEM with electron energy-loss spectroscopy (EELS) to resolve how oxidation effects distort the local structure. Fig. 4(e) shows a representative HR-TEM image in which both crystalline and amorphous features (near the edges) are visible. We measured EELS spectra from the two positions marked with black and red dots, corresponding to a well-crystallized region and a disordered region, respectively. Analysis of the EELS spectra confirms that there are only minor traces of oxidation due to the native oxide layer that are more pronounced near the edges, i.e., in the amorphous region (Fig. 4(f)).

TEM analysis demonstrates the crystalline nature of the platelets, with the presence of low-angle grain boundaries suggesting bending at the single-platelet scale. Furthermore, EELS analysis indicates that amorphization and oxidation primarily affect the edges of the sample.

## IV. DISCUSSION

The native 2D platelet structures quantified by SEM reveal that Si-4H from HP synthesis has a far more complex hierarchical structure than was previously understood. Our size analysis quantifies the 1-μm by 10-nm sizes of the pristine quasi-2D nano-platelets. In comparison, our previous indirect analysis via XRD Rietveld refinement and FFT-TEM analysis concluded that flake-like crystallites (i.e. *nano-flakes*) were formed from a high density of stacking faults [28]. In this work we use multi-scale microscopy to perform direct imaging of Si-4H nanostructure;cthe dimensions and stacking of the platelets we observe differ from the previous view of highly-oriented nano-flakes. The loose packing between the platelets indicates that the origin of the nanostructure in Si-4H may be strongly linked with specific aspects of the synthesis and crystal growth, e.g., transition mechanism geometry, rather than being solely due to the presence of stacking faults. Beyond presenting questions as to Si-4H growth mechanism, the platelets also demonstrate another key length scale for Si-4H, that might play a role in dictating its optoelectronic properties.

So far, two synthetic approaches have demonstrated hexagonal Si-4H. The crystallization of Si-4H from Si$_{24}$ annealing has shown a high degree of preferred orientation [45]; it is possible that an orientation relationship exists also between Si-4H and the Si-III precursor, which might favour selective growth of the platelets depeding on their orientation. Precise characterization of the Si-III→Si-4H transition mechanism to corroborate this hypothesis is, however, beyond the scope of the present study. Polycrystalline Si-4H (with ~0.5 μm crystalline domains) formed via annealing of metastable single-crystal Si$_{24}$ exhibit a ~1.2 eV absorption onset [45]. Our HP synthesis resulted in nanostructured Si-4H, and spectroscopy measurements suggest a bandgap of ~1.6 eV [29], demonstrating the influence of grain size and microstructure on Si-4H's optoelectronic properties. The additional structural complexity in our hierarchical Si-4H offers novel opportunities for bangdap engineering, and future studies could further optimize Si-4H absorption properties by acting on the synthesis procedure and the crystal size. The availability of hexagonal Si with different microstructure and absorption efficiency could open the way for the development of novel Si-based multi-junction solar cells, in which the different textures would enhance absorption of different regions of the solar spectrum.

Our study reveals two key features of Si-4H that were previously not identified. We demonstrate an intrinsic nanostructure, as the sample is composed by quasi-2D grains with wide size distribution. The stronger confinement of the platelets along one dimension, with estimated thickness of tens of nm, may have an impact on the bandgap value, and thus the absorption efficiency, of Si-4H particles. Interestingly, the weak inter-platelet binding enables separation and dispersion of the particles without affecting their crystal quality, and could thus be used for nano-patterning novel devices or to deposit thin absorbing layers for photovoltaic panels.

Our work also demonstrate low-angle grain boundaries, which suggest an inherent malleability at the scale of single platelets. The grain boundaries developed upon bending demonstrate a ductility in the platelets that can withstand mechanical stresses by changing shape instead of fracturing, despite being only ~10s nm thick. The mechanical properties of Si-4H platelets are of interest also for their relevance to strain-engineering, as hexagonal Si is predicted to become a direct-bandgap material when strained in its bullk form [34] or at the atomic scale, e.g., via Ge substitution [18]. The ductility of Si-4H platelets could enable the realization via strain-engineering of the first direct-bandgap Si material, a key deveolpment for the future of solar industry.

## V. CONCLUSION

In this work, we characterize the multi-scale structure of Si-4H at all the relevant length-scales to provide the baseline for future applications-focused studies into optoelectronic materials design. We demonstrate how the macroscopic pellets generated at high pressures are comprised entirely of large collections of sub-μm quasi-2D *platelets* that pack into bundles. The loose interaction forces and preferential packing along the wider faces enabled us to disperse the platelets to study their shape and local structure in detail, showing that they have sizes of ~1000 ± 400 nm in diameter with only <100 nm thicknesses. With atomic resolution, we revealed high crystallinity with minimal oxidation in each platelet, and demonstrated how two offset domains suggest that the platelet can bend by forming a low-angle grain boundary.

With our multi-scale structural characterization, we demonstrated 2D Si-4H platelets that present opportunities for novel applications of hexagonal Si, a potential photovoltaic candidate. The successful dispersion procedure allows us to access and manipulate quasi-single platelets and it could be applied in the design and patterning of new optoelectronic and solar devices. Besides the practical advantages provided by the platelets' lamellar structure, their dimensions could also present opportunities to optimize Si optoelectronic performance, as both quantum confinement and oxidation modify Si-4H's optoelectronic properties. Furthermore, the flexibility at the single-crystal scale suggests that Si-4H platelets could accommodate, without affecting the crystal quality nor fracturing, the strain required to transform thexagonal Si into a direct bandgap material.

**Acknowledgments:** The authors thank Dr. Kristina Spektor and Dr. Wilson Crichton (ESRF, Grenoble, France) and Paraskevas Parisiadis and Louis Amand (IMPMC, Paris, France) for the help with Si-4H synthesis. Initial XRD characterization was performed in collaboration with Dr. Benoit Baptiste, Ludovic Delbes (IMPMC) and Jon Lee (LLNL). SZ was supported by the US Office of Naval Research under Grant No. N00014-17-1-2283. *Work at the Molecular Foundry was supported by the Office of Science, Office of Basic Energy Sciences, of the U.S. Department of Energy under Contract No. DE-AC02-05CH11231.*
. S.P.'s PhD fellowship has been provided by ED397, and part of S.P.'s work was financed by LLNL (Livermore, CA) through the JHEDS Summer Intern program. Contributions from LEDM were performed under the auspices of the U.S. Department of Energy by Lawrence Livermore National Laboratory under Contract DE-AC52-07NA27344, with additional support from the Lawrence Fellowship.